\documentclass{article}

\usepackage{authblk}
\usepackage{mathptmx}       % selects Times Roman as basic font
\usepackage{helvet}         % selects Helvetica as sans-serif font
\usepackage{courier}        % selects Courier as typewriter font
\usepackage{type1cm}        % activate if the above 3 fonts are
\usepackage{makeidx}         % allows index generation
\usepackage{graphicx}        % standard LaTeX graphics tool
\usepackage{multicol}        % used for the two-column index
\usepackage[bottom]{footmisc}% places footnotes at page bottom

\makeindex             % used for the subject index
\begin{document}

\title{Quantum interference effects in electron transport: How to select suitable molecules for logic gates and thermoelectric devices}
%\titlerunning{Quantum interference effects in electron transport}
\author{Robert Stadler}
\affil{University of Vienna, Department of Physical Chemistry\\ Sensengasse 8/7, A-1090 Vienna, Austria \\ Email: robert.stadler@univie.ac.at}
\maketitle

\abstract{Since the concepts for the implementation of data storage and logic gates used in conventional electronics cannot be simply downscaled to the level of single molecule devices, new architectural paradigms are needed, where quantum interference (QI) effects are likely to provide an useful starting point. In order to be able to use QI for design purposes in single molecule electronics, the relation between their occurrence and molecular structure has to be understood at such a level that simple guidelines for electrical engineering can be established. We made a big step towards this aim by developing a graphical scheme that allows for the prediction of the occurrence or absence of QI induced minima in the transmission function and the derivation of this method will form the center piece of this review article. In addition the possible usefulness of QI effects for thermoelectric devices is addressed, where the peak shape around a transmission minimum is of crucial importance and different rules for selecting suitable molecules have to be found.}

\section{Introduction}
\label{sec:1}
In the vivid field of molecular electronics, where tremendous advances in the theoretical description as well as experimental characterization of the conductance of single molecule junctions have been achieved in the last two decades, two concepts emerged recently which received a lot of attention: i) devices based on quantum interference (QI) effects~\cite{Baer2002,Stadler2003,StadlerNanotech2004,DijkOrgLett2006,Andrews2008,MarkussenJCP2010} and ii) thermoelectric applications (TA)~\cite{Bergfield2009,Finch2009,Bergfield2010,Nozaki2010,KamalMarkussen2011}. QI based devices are utilizing the wave nature of electrons and the resulting possibility for destructive interference of transmission amplitudes for the design of data storage elements~\cite{Stadler2003}, transistors~\cite{Baer2002,Andrews2008} or logical gates~\cite{StadlerNanotech2004}. TAs on the other hand focus on the conversion of thermal gradients to electric fields for power generation or vice versa for cooling or heating with so called Peltier elements, an application for which not only homogeneous bulk materials but also single molecule junctions might be used. The efficiency of a thermoelectric device is measured by a dimensionless number, the figure of merit $ZT$, which can be increased by steep declines in the transmission function describing the electronic conductance of a junction~\cite{Finch2009,Paulsson2003}. This latter requirement for a high ZT value provides a link between QI and TA in the sense that QI effects can cause the transmission function to change from rather high to rather low values or vice versa within a small energy range close to the Fermi level of the electrodes for some molecules.

In this article I review my personal involvement in studying QI effects in single molecule electronics, where I start with introducing QI based concepts for data storage~\cite{Stadler2003} and logical gates~\cite{StadlerNanotech2004}, then highlight a graphical design scheme for predicting the occurence of QI induced minima in the transmission function~\cite{MarkussenNanoLett2010} and the energies at which they occur~\cite{MarkussenPCCP} in their dependence on the molecular structure. In the second half of the paper I present a method to analyze the structure dependence of the asymmetry of interference dips - which is crucial for TA - from simple two site tight-binding models, where one site corresponds to a molecular $\pi$ orbital of the wire and the other to an atomic $p_z$ orbital of a side group, which allows for an analytical characterization of the peak shape in terms of just two parameters~\cite{StadlerFano}.

\section{Mono-molecular data storage and logic gates}
\label{sec:2}

\begin{figure}
\includegraphics[scale=.67]{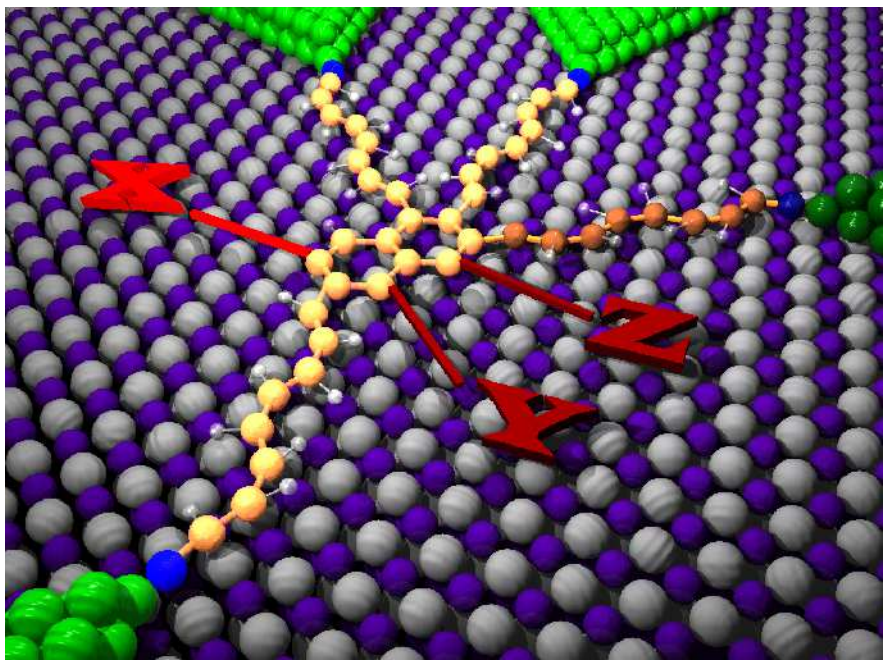}
\caption{A molecular data storage scheme based on an aromatic molecule (naphthalene) bonded to four gold electrodes by sulphur atoms and poly-acetylene wires. For the surface an insulator has to be chosen to prevent cross-talk between the electrodes. The variables X, Y and Z could either be chemical substituents or alternatively connections to further electrodes. Reprinted with permission from Ref.~\cite{Stadler2003}, Copyright (2003), Institute of Physics.}
\label{fig:1}
\end{figure}

An important task in molecular electronics is to implement active device behaviour in an electric circuit by using the structural or electronic properties of individual molecules. Since the final aim is to arrive at an overall computer architecture which is smaller than currently available semiconductor technology~\cite{StadlerArch,StadlerC601,StadlerC602}, it should be kept in mind that the molecules are connected to metal electrodes which need to be wired to the outside world for being addressable. Even if feasible, it would be highly impractical to consider this metal wiring scaled down to an atomic level due to the introduction of additional wire dependent electron scattering and the necessity for those wires to be then absolutely defect free. One way to avoid these problems can be found in exploring the possibilities for intra-molecular circuits, where the device behaviour of one single molecule mimics not only a single switch but e.g. a two-input logic gate or even a more complex part of a logic circuit.

\begin{figure}
\includegraphics[scale=0.54]{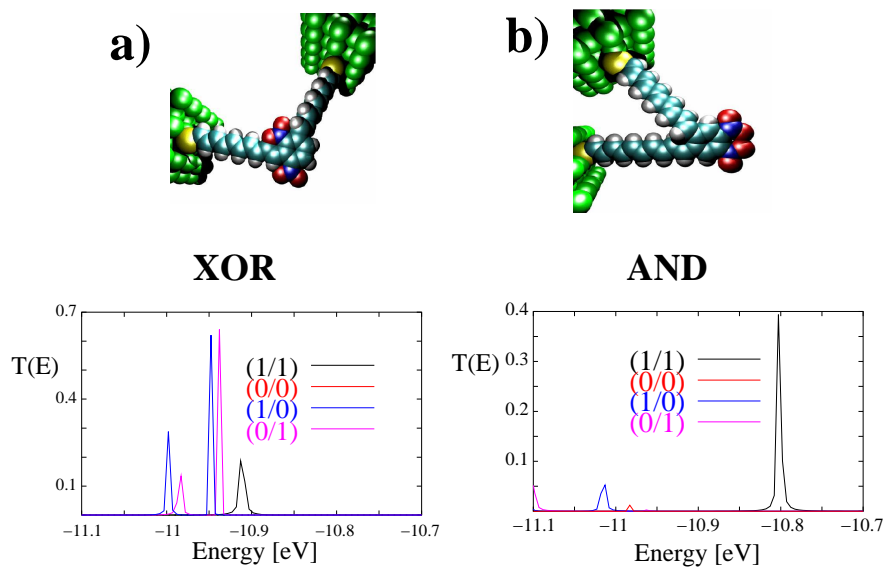}
\caption{A benzene molecule is connected to two gold electrodes by sulfur atoms and polyacetylene wires and the transmission through the molecule can be designed to have the characteristics of either a) a XOR-gate or b) an AND-gate. The type of gate is defined by the positions on the benzene ring, where the wires are connected, namely in a meta-configuration for the XOR-gate and in an ortho-configuration for the AND-gate. It is also crucial on which ring positions the nitro-groups are bonded to the molecules. It is envisioned that they are rotated for defining the input (with unspecified means for inducing this rotation in praxis), where the substituents interact with the $\pi$ electrons of the benzene ring, when they are planar to the ring and there is no interaction when their rotational state is perpendicular to the benzene molecule. This means that for an input of (1/1) both nitro-groups are in the same plane as benzene, for (1/0) and (0/1), one nitro-group is planar the other rotated by 90$^{\circ}$, respectively, and for (0/0) both groups are perpendicular. The output is then defined by the electron transmission probability through the junction, with a high transmission representing '1' and a low transmission representing '0'. The figure shows that the correlation between input and output for the respective molecular junctions fulfills the truth tables for XOR and AND gates. Reprinted with permission from Ref.~\cite{StadlerNanotech2004}, Copyright (2004), Institute of Physics.}
\label{fig:2} 
\end{figure}

The key idea for the design of such intra-molecular circuits first presented in Ref.~\cite{Stadler2003} and illustrated in Fig.~\ref{fig:1} is to take into account the influence that symmetry has on electron transmission through highly delocalised $\pi$ states of aromatic molecules, which are connected to a reference electrode and a given number of output electrodes. It is assumed that the electron transfer process across the molecule could in principle be tuned by varying chemical substituents, e.g. with the help of redox reactions, or by locally acting on them, e.g. by using an STM tip. This would control the output current by changing the interference pattern of the electron scattering process, and the chemical substitution pattern could be used for data storage. Such an architectural scheme, although its structure is defined in the space domain, would share some favourable characteristic features with quantum computing, such as the possibility for dense storage scaling as 2$_N$, where N is a parameter reflecting the size of the system. While in Ref.~\cite{Stadler2003} this original idea has been first proposed, the concept has been extended in Ref.~\cite{StadlerNanotech2004} for the design of small logic circuits, where first electron transport calculations taking into account the detailed three-dimensional structure of the molecular orbitals of benzene molecules and using NO$_2$ groups as ligands on an extended H\"{u}ckel level have been performed (Fig.~\ref{fig:2}). 

Since it is not possible in the case of intra-molecular architectures to decompose them into single gates and wires~\cite{StadlerTB1,StadlerTB2} as is done in semiconductor industry, architectural design needs to take into account the system as a whole. The overall aim is to find a molecular structure on which a maximum amount of computational functionality for information storage and/or processing can be imposed. For this task a simple graphical method can be employed~\cite{StadlerNanotech2004,MarkussenNanoLett2010}, which does not take into account the full complexity of the electron transport through the molecule. This method is derived in Sec.~\ref{sec:3} from a topological tight binding (TB) description, where each atom contributing to the $\pi$ system is only described by a single atomic orbital (AO). The use of such a description for the comparison of different molecular topologies consisting of rings or chains of AOs already allows for an assessment of the applicability of particular molecular structures for architectural design, since its predictions have been validated by density functional theory (DFT) calculations in Ref.~\cite{MarkussenNanoLett2010} as pointed out in more detail in Sec.~\ref{sec:3}. 

\section{A graphical design scheme and its validation by density functional theory}
\label{sec:3}
Within a single-particle picture, the transmission probability of an electron entering a molecular junction with an energy $E$ can be reduced to
\begin{equation}
\mathcal{T}(E) = \gamma(E)^2|G_{1N}(E)|^2
\end{equation}
assuming that the Hamiltonian describing the molecule $H_{mol}$ is given in terms of a basis consisting of localized atomic-like orbitals, $\phi_1,\phi_2,\ldots,\phi_N$, and that only the two orbitals $\phi_1$ and $\phi_N$ couple to the leads~\cite{MarkussenNanoLett2010,MarkussenPCCP}.

Often the energy dependence of the lead coupling strength, $\gamma$, can be
neglected. It then follows that the transport properties are entirely governed by
the matrix element $G_{1N}(E)$. The latter can be obtained using Cramer's
rule
\begin{equation}
\label{eqn.g1n}
G_{1N}(E)=\frac{det_{1N}(E-H_{mol})}{det(E-H_{mol}-\Sigma_L-\Sigma_R)}
\end{equation}
where $det_{1N}(E-H_{mol})$ is the determinant of the matrix obtained
by removing the 1st row and $N$th column from $E-H_{mol}$ and
multiplying it by $(-1)^{1+N}$. Taking the Fermi energy to be zero without loss of
generality, we can then state the condition for complete destructive interference of the zero bias conductance, $G(E_F)=0$, as
\begin{equation}\label{eqn.destructive}
det_{1N}(H_{mol}) = 0.
\end{equation}

\begin{figure}
\includegraphics[scale=0.48]{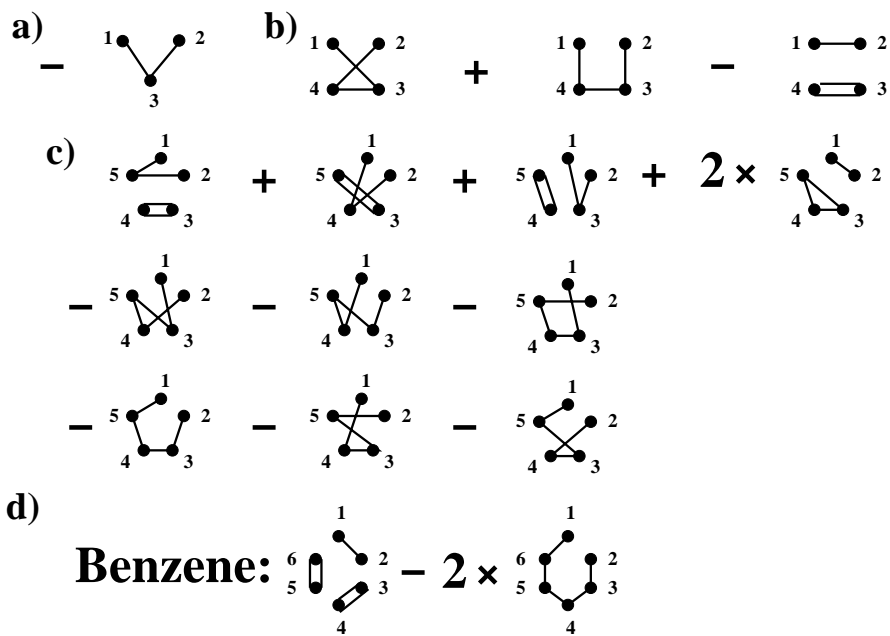}
\caption{Visualisation of the terms in $det_{1N}(H_{mol})$ (see text) for electron transport between orbitals $1$ and $2$ for a) three, b) four and c) five atomic orbitals, respectively. Panel d) is for six orbitals, with the coupling pattern of benzene being taken as a boundary condition. Atomic orbitals are drawn as dots; lines connect two orbitals when there is a factor in the respective term which has both of their indices and double lines indicate that this factor is squared. Reprinted with permission from Ref.~\cite{StadlerNanotech2004}, Copyright (2004), Institute of Physics.}
\label{fig:3}       
\end{figure}

For a general three site system, where the coupling constants between the orbitals $a_{12}$, $a_{13}$ and $a_{23}$ are unspecified and their onsite energies $\varepsilon_i=0$ for all three atomic orbitals, the matrix $H_{mol}$ can be written as
\begin{eqnarray}
\label{eqn.mat}
H_{mol}=\left[ \begin{array}{ccc}
-E & a_{12} & a_{13} \\
a_{12} & -E & a_{23} \\
a_{13} & a_{23} & -E \\
\end{array}\right],\ \
\end{eqnarray}

where for $E=E_F$
\begin{eqnarray}
det_{1N}(H_{mol}) = - a_{13} a_{23}.
\end{eqnarray}
For general four and five site systems under the same conditions this expression is replaced by 
\begin{eqnarray}
det_{1N}(H_{mol}) = a_{13} a_{24}  a_{34} + a_{14}  a_{23}  a_{34} - a_{12}  a_{34}^2
\end{eqnarray}
and 
\begin{eqnarray}
det_{1N}(H_{mol}) = a_{34}^2  a_{15} \  a_{25} + a_{35}^2  a_{14}  a_{24} + a_{45}^2  a_{13}  a_{23} + 2  a_{12}  a_{34}  a_{35}  a_{45} \\ - a_{13}   a_{24}  a_{35}  a_{45} - a_{14}  a_{23}  a_{35}  a_{45} - a_{13}  a_{25}  a_{34}  a_{45} - a_{15}  a_{23}  a_{34}  a_{45} - a_{14}  a_{25}  a_{34}  a_{35} \nonumber \\ - a_{15}  a_{24}   a_{34}  a_{35}, \nonumber            
\end{eqnarray}
respectively, where $det_{1N}(H_{mol})$ is visualised as graphs in Figs.~\ref{fig:3}a, b and c for three, four and five orbitals respectively. Each term in the sums is represented by a different drawing of the orbitals as dots, with lines connecting them if there is a factor in the term which has both of their indices, and with double lines when the factor is squared. Looking at the structure of these graphs it can be seen that the terms represent all possibilities for a path between orbitals $1$ and $2$, where all orbitals of the molecule have to be either traversed within the path or within a closed loop, which for the comparatively small number of orbitals shown in Fig.~\ref{fig:3} are either double lines or triangles (~\ref{fig:3}c) but can be larger loops for a larger structure. A larger number of orbitals leads also to an increase in the number of possible paths. But the individual terms contain more couplings as well and if only one bond is missing, then the path does not exist. For six orbitals, there are 46 possible paths, but for the bonding pattern of a particular molecule this number considerably reduces and becomes two for benzene (Fig.~\ref{fig:3}d) with the electrodes attached in an ortho configuration.

\begin{figure}
\includegraphics[scale=0.55]{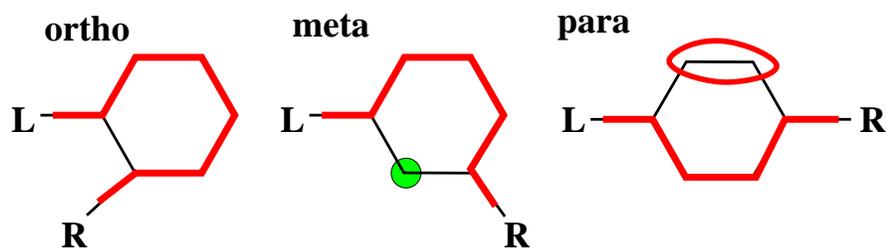}
\caption{Demonstration of how the graphical QI prediction scheme is applied for electron transport through benzene in ortho, meta and para connections. For the connections without QI, a continuous path can be drawn either with (para) or without (ortho) pairing up remaining sites. This is not possible for the only connection, which exhibits QI for benzene (meta), where the isolated site has been marked by a green spot. Reprinted with permission from Ref.~\cite{MarkussenNanoLett2010}, Copyright (2010), American Chemical Society.}
\label{fig:4}       
\end{figure}

The rationale for applying the graphical scheme for the prediction of QI effects is based on the fact that viable paths must exist for $det_{1N}(H_{mol})$ to be finite, which is a precondition for $G(E_F)>0$ (see Eqns.~\ref{eqn.g1n} and~\ref{eqn.destructive}). A path - in this interpretation one of the terms in the sum defining $det_{1N}(H_{mol})$ - is viable only if there are bonds between atomic orbitals corresponding to all factors in this term. As can be extrapolated from the examples in Fig.~\ref{fig:3} the general rule is that all atomic sites have to be either part of a continuous line connecting the two sites coupled to the leads (1 and 2 in Fig.~\ref{fig:3}) or be part of a closed loop (in praxis i.e. for realistic molecules only pairs - with a bond between them - of isolated atomic orbitals occur, triangles or even larger loops are uncommon). In Fig.~\ref{fig:4} we show how the method is applied for predicting QI for the well-known case of benzene connected to two leads in ortho, meta and para configurations. While for ortho all atomic sites are part of a continuous line from lead to lead, for para the two orbitals not on this line form a pair with a bond between them (marked by a red elliptical confinement in Fig.~\ref{fig:4}). As is well known these two configurations do not exhibit QI. The meta setup on the other side has an isolated atomic site without a neighbour which is not part of the curve (highlighted by a green spot in Fig.~\ref{fig:4}) and therefore QI results from the scheme. While in more complicated molecules there may be several ways to draw continuous lines between the leads, the crucial point is whether it is possible to draw it in a way that the general rule named above is fulfilled. If for instance the curve for meta would be drawn in the other direction, then three sites would be left on the other side of the molecule, where two of them could form a bonded pair but again there would be one orbital remaining.

\begin{figure}  
\includegraphics[scale=0.31]{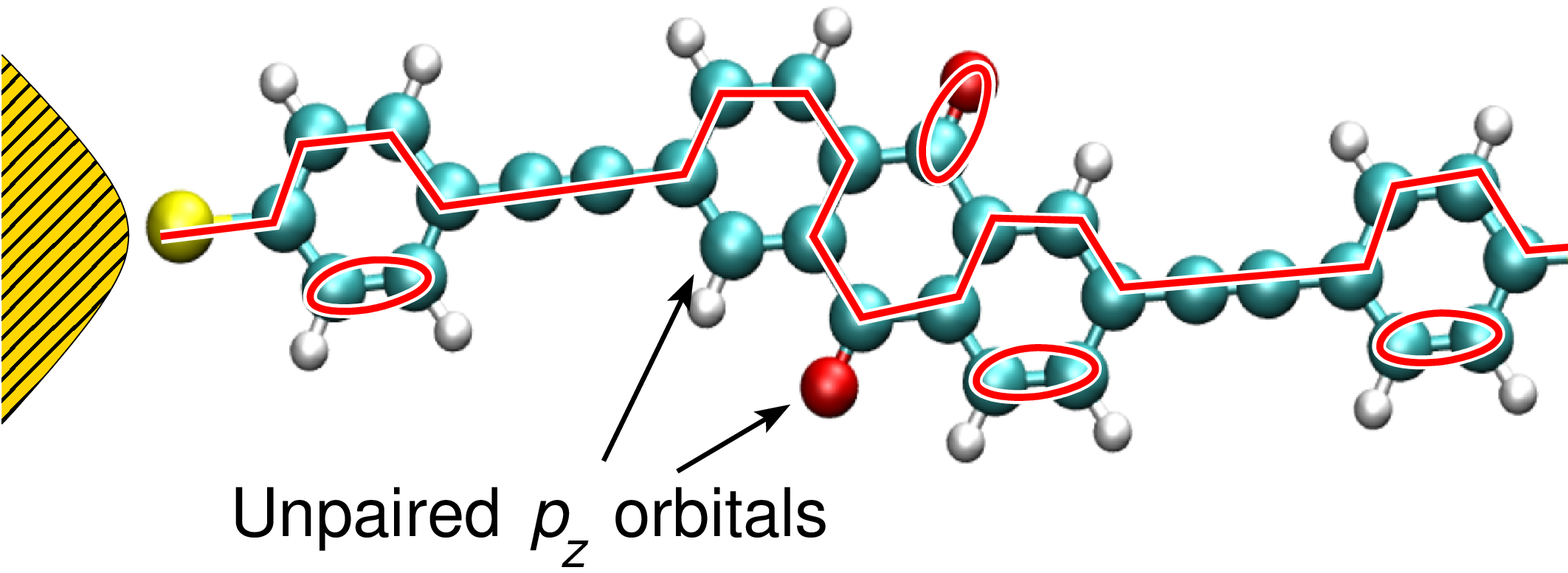}
\caption{Illustration of the graphical scheme (left) for predicting a QI induced minimum in the electron transmission function (right) of an anthraquinone molecule between gold electrodes. Reprinted with permission from Ref.~\cite{MarkussenNanoLett2010}, Copyright (2010), American Chemical Society.}
\label{fig:5}       
\end{figure}

\begin{table}
\caption{DFT conductance values in units of $G_0=2e^2/h$ for a range of anthraquinone molecules where some experience QI effects (I1-I5) and others do not (N1-N5). Note that the conductance values in the left column are systematically lower than those in the right column. This is a result of destructive QI occurring close to the Fermi energy for the molecules I1-I5. Reprinted with permission from Ref.~\cite{MarkussenNanoLett2010}, Copyright (2010), American Chemical Society.}\label{tab.cond}
\begin{tabular}{p{2.4cm}p{3.2cm}p{2.4cm}p{3.2cm}}
\hline\noalign{\smallskip}
  Molecule       & $G$ ($G_0$) & Molecule & $G$ ($G_0$) \\
\noalign{\smallskip}\hline\noalign{\smallskip}
I1 &   7.71 10$^{-8}$    &   N1    &   1.38 10$^{-2}$   \\
I2 &   1.82 10$^{-8}$    &   N2    &   8.57 10$^{-2}$   \\
I3 &   3.03 10$^{-4}$    &   N3    &   7.09 10$^{-3}$   \\
I4 &   3.55 10$^{-4}$    &   N4    &   8.98 10$^{-3}$   \\
I5 &   8.84 10$^{-4}$    &   N5    &   6.19 10$^{-3}$   \\
\noalign{\smallskip}\hline\noalign{\smallskip}
\end{tabular}
\end{table}

In Ref.~\cite{MarkussenNanoLett2010} we studied ten junctions with anthraquinone molecules (Fig.~\ref{fig:5}), which only differed in the position of the two oxygen side groups on the molecules, and categorized them into two groups, where five of them exhibited QI inside the energy range of the HOMO-LUMO gap (I1-I5) and the others (N1-N5) did not. We made this distinction on the basis of our graphical scheme and our conclusions were in good agreement with DFT calculations, where as can be seen from Table~\ref{tab.cond} the conductance of molecules I1-I5 is systematically and significantly lower than the conductance of molecules N1-N5. This prediction, however, was based on the assumption that the onsite energy of the side group would be approximately equal to the onsite energy of the carbon $p_z$ orbitals. From the side group analysis in Ref.~\cite{MarkussenPCCP} it was found that this assumption is questionable for oxygen with a side group energy of -2 eV (relative to the Fermi level). In the same article we showed that more quantitative estimates of the transmission node position can be obtained from a straightforward generalization of the graphical scheme to the case of finite (and varying) on-site energies. This scheme was then used to analyze the transmission nodes in linear and aromatic molecules with side groups. For linear molecular chains a single transmission node occurs at an energy corresponding to the energy of the side group $\pi$-orbital, while for aromatic molecules the nodal structure of the transmission function is in general more complex due to a non-trivial interplay between molecular topology and side-group onsite energy, which explains the success of the original scheme for categorizing the anthraquinone molecules as exhibited in Table~\ref{tab.cond} even beyond the limiting conditions assumed for its derivation~\cite{MarkussenPCCP}.

\section{Transmission peak-shape engineering for thermoelectric devices}
\label{sec:4}

\begin{figure}                                                                                           \includegraphics[scale=0.72]{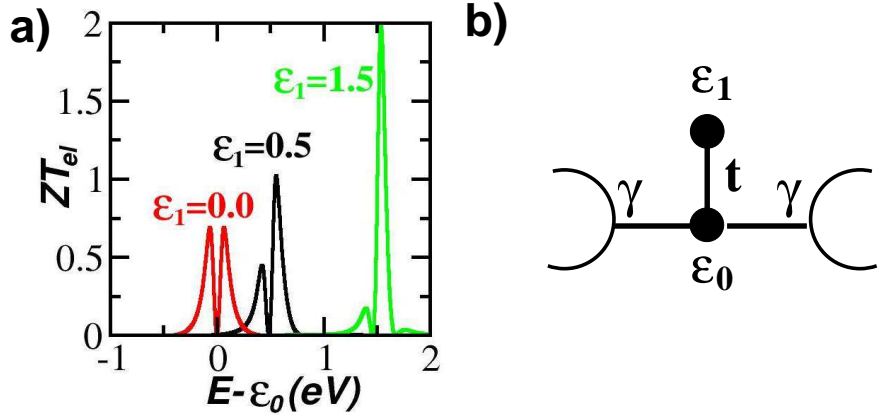}
\caption{(a) Analytical TB results for the purely electronic thermoelectric figure of merit $ZT_{el}$ calculated from the transmission functions of a two-site model (b) according to Eqns. (8) and (11) with $\gamma = 0.5\,$eV, $\varepsilon_0=0.0\,$eV and $|t|=0.5\,$eV for $\varepsilon_1=0.0\,$eV (red), $0.5\,$eV (black) and $1.5\,$eV (green). Reprinted with permission from Ref.~\cite{StadlerFano}, Copyright (2011), American Institute of Physics.}
\label{fig:6}   
\end{figure}

The TB-model, which was first introduced in Ref.~\cite{StadlerFano} aims at a separation of the structural aspects governing the key quantity for the electronic contribution to thermoelectric properties. I am referring to the peak shape of interference dips in the transmission function, for which analytical expressions can be derived in dependence on just two parameters given a few simplifying assumptions are made. First, as for the graphical scheme in the last section, it is assumed that only the $\pi$ electrons are relevant in the energy range of interest, which allows us to describe molecules within a TB model with only one $p_z$ AO for each carbon- or hetero-atom. In a second step, we look at MOs of t-stub wires~\cite{Porod1992,Porod1993} without the side group attached, which can be obtained by diagonalizing the TB-Hamiltonian for a chain of carbon atoms with a certain length and from them pick the single MO which is closest in energy to the $p_z$ AO or fragment orbital of the side group. If we now assume a two-site single-particle picture where the transmission function for electrons incident on a molecular junction with an energy $E$ is dominated by this MO and its interaction with the side group AO, it can be written as
\begin{equation}
\mathcal{T}(E)= \frac{4\gamma^2}{(E-\varepsilon_0-\frac{t^2}{E-\varepsilon_1})^2+4\gamma^2}, 
\label{Eq:two-site-trans}
\end{equation}
as derived in Ref.~\cite{StadlerFano}, with $\gamma$ being the lead coupling within a wide band approximation, $\varepsilon_0$ the wire MO energy, $\varepsilon_1$ the side group AO energy and $t$ the coupling between them. For an illustration see Fig.~\ref{fig:6} (b).
Since the thermoelectric properties of a molecular junction are optimized if the transmission function rapidly change from a minimum to a maximum, we can define the energetic difference $\Delta E$ between the transmission minimum and its closest maximum as a simple criterion. It follows from Eq. \ref{Eq:two-site-trans} that the transmission is zero at energy $E=E_0=\varepsilon_1$,~\cite{MarkussenPCCP} while for determining the energy of transmission peaks with $\mathcal{T}=1$ the quadratic equation $E-\frac{t^2}{E-\Delta\varepsilon}=0$ has to be solved where we made the substitution $\Delta\varepsilon=\varepsilon_1-\varepsilon_0$ and set $\varepsilon_0=0$ without loss of generality for our following arguments. The solutions to this equation $E_{1,2}$ and the norm of their differences with the energy of the transmission zero $E_0$ can then be obtained as
\begin{eqnarray}
E_{1,2}=\frac{\Delta\varepsilon}{2}\pm\frac{1}{2}\sqrt{\Delta\varepsilon^2+4t^2} &  \\
|E_{1,2}-E_0|=|-\frac{\Delta\varepsilon}{2}\pm\frac{1}{2} \sqrt{\Delta\varepsilon^2+4t^2}|,
\end{eqnarray}
where it can be seen that for finite $\Delta\varepsilon$ there will always be one solution for $|E_{1,2}-E_0|$ which will be closer to zero than the other. In other words only under this condition will the energy difference of the two transmission maxima from the transmission minimum be asymmetric for the two-site model, while for $\Delta\varepsilon=0$ we get $|E_{1,2}-E_0|=\pm t$ and the peak shape around the minimum in $\mathcal{T}(E)$ will therefore be symmetric. In the following I will refer to the smaller of the two solutions for $|E_{1,2}-E_0|$ as $\Delta E$, which is used as a measure indicative of thermoelectric efficiency where the thermopower is expected to be high for small $\Delta E$. An important conclusion at this point is that $|E_{1,2}-E_0|$ is completely independent of the lead-coupling $\gamma$. Rapid transmission changes being independent of lead-coupling broadenings is a general advantage of utilizing QI since both transmission peaks and transmission minimums are defined entirely by the molecular topology. As a consequence $\Delta E$ is completely defined by just two parameters, where peak shapes of the transmission function for molecular t-stubs are found to be the more asymmetric, the larger $\Delta\varepsilon$ and the smaller $t$. We note that these two parameters correspond to the length and height of the tunnelling barrier introduced in Refs.~\cite{Porod1992,Porod1993} for a separation of the sidearm in mesoscopic waveguide t-stubs, where the authors also found that distinctly asymmetric line shapes only start to appear for weak coupling and large barrier heights.

The efficiency of a thermoelectric material can be characterized by the dimensionless figure of merit, $ZT$, given by \begin{equation}
ZT = \frac{S^2G\,T}{\kappa_{ph}+\kappa_{el}} \label{ZTdef},\\
\end{equation}
where $S$ is the Seebeck coefficient, $G$ the electronic conductance, $T$ the temperature, and $\kappa_{ph}$ and $\kappa_{el}$ are the phonon- and electron contributions to the thermal conductance, respectively~\cite{StadlerFano}. When the phonon contribution is neglected ($\kappa_{ph}=0$) one obtains the purely electronic $ZT_{el}$, which is illustrated in Fig.~\ref{fig:6} (a) for different values of $\Delta\varepsilon$. Clearly the maximum $ZT_{el}$ values increase when the transmission function becomes more asymmetric which can be brought about by a rise in $\Delta\varepsilon$ for a constant $t$. While very large $ZT_{el}$ values have been predicted for symmetric transmission functions\cite{Bergfield2010}, the results in Fig. \ref{fig:6} (a) show that asymmetric transmission functions are even more promising for a thermoelectric purpose. 

\begin{figure}                                                                                           \includegraphics[scale=0.48]{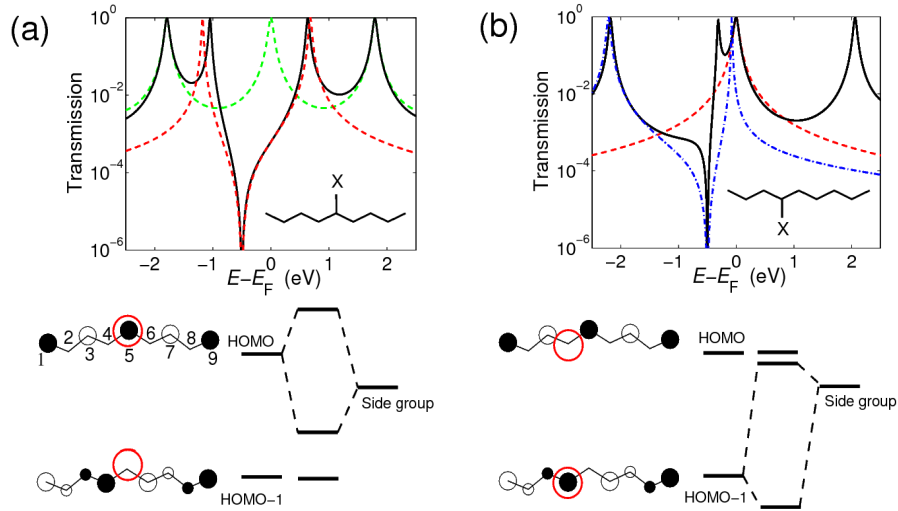}
\caption{Transmission functions (upper panels) and MO-AO hybridisation schemes (lower panels) for (a) C9-5 and (b) C9-4 type molecular stubs with a side group AO X ($\varepsilon_{sg}$=-0.5 eV), where the respective AO topologies are given as insets in the transmission figures. The labels 5 and 4 correspond to the atom number in the carbon chain to which the side group is coupled, where we indicate the numbering we use for the atoms in the carbon chain in the lower part of panel (a). The line types for $\mathcal{T}(E)$ correspond to full TB AO calculations with (black) and without (green) the side group, and a 2-site MO model with the side group AO and the HOMO ($\varepsilon_{HOMO}$=0 eV, dashed red) or the HOMO-1 ($\varepsilon_{HOMO-1}$=-1.8 eV, dashed blue), respectively. The Fermi level E$_F$ is assumed to be identical to $\varepsilon_{HOMO}$ of C9 (without any side groups). The lower panels also show the topologies of the HOMO and HOMO-1 where the respective position of the side group is indicated by red empty spheres, where depending on $|\Delta\varepsilon|$ and $|t_{sg}|$ different hybridization patterns are observed. Reprinted with permission from Ref.~\cite{StadlerFano}, Copyright (2011), American Institute of Physics.}
\label{fig:7}  
\end{figure}

In order for our two-site model to be useful for the analysis of t-stub molecular wires we need to be able to deduce the parameters $t$ and $\Delta\varepsilon$ from the full topology of the wires' $p_z$ AOs including the side group, which does not always have to be another carbon-site. For this we start from an AO-TB-model where the onsite energy of each $p_z$ orbital of the carbon chain is set to zero and the couplings between neighbouring sites are defined as -2.9 eV. From a diagonalization of the chain \textit{without} the side group, where I refer to Ref.~\cite{StadlerFano} for the full mathematical details, its MO energies and the orbital weight on each chain site for each MO can be obtained. Within this scheme the side group AO energy $\varepsilon_1$ and an initial value for the MO-AO coupling $t$ can be freely varied so that it reflects the side groups chemical nature. This initial value for $t$ is then multiplied by the orbital weight at the chain site to which the side group is bonded. The relevant MO is typically the one closest in energy to $\varepsilon_1$. The two site model finally contains this latter MO (with energy $\varepsilon_0$) and the side group AO, where the lead coupling $\gamma$ in the two-site model is derived from the wide band lead coupling $\Gamma$ used in the full AO-TB model multiplied by the squared orbital weight on the terminal sites in the chain.

In Fig.~\ref{fig:7} it is illustrated how this works in practice and which kind of information can be obtained from such an analysis. The figure focuses on molecular wires with nine atoms in the chain (C9), where a side group with $\varepsilon_1=-0.5\,$eV is attached to the fifth (C9-5, Fig.~\ref{fig:7}a) and the fourth (C9-4, Fig.~\ref{fig:7}b) site of the chain. The solid black lines in the transmission functions in the upper panels of Fig.~\ref{fig:7} are the results of full TB AO calculations including the side group, while the dashed red and blue lines come from 2-site TB MO models considering only the side group AO and the HOMO (red) or HOMO-1 (blue) of the unsubstituted chain, respectively. The green line in Fig.~\ref{fig:7}a shows the TB AO result without the side group, where it can be clearly seen that the peak in $\mathcal{T}(E)$ corresponding to resonance through the HOMO splits into two, when the HOMO hybridizes with the side group AO, whereas the other two visible peaks caused by the HOMO-1 and LUMO remain unaffected. The situation is quite different for C9-4 (Fig.~\ref{fig:7}b), which exhibits a distinctly asymmetric QI feature and the red line does not reproduce the black one at all and instead represents a single site Lorenz peak at the HOMO energy. The blue line on the other hand shows very good correspondence with the TB AO transmission, which means that the side group AO interacts mainly with the HOMO-1 and not with the HOMO.
This difference between the two molecular t-stubs can be understood in terms of the topologies of the MOs plotted in the lower panels of Fig.~\ref{fig:7}, where the HOMO has a high weight on site 5 but none on site 4 and in contrast the HOMO-1 has its largest weight on site 4 and none on site 5. 

\begin{table}
\caption{2-site MO model parameters $|\Delta\varepsilon|$ and $|t_{sg}|$ for a range of molecular stubs of varying length and chemical nature of side groups. The energy of the transmission zero $E_0$ and its difference to that of the nearest transmission maximum $|\Delta E|$ are evaluated from transmission functions calculated from DFT. The maximum values for the thermopower $S$ and the electronic contribution to the figure of merit $ZT_{el}(E)$ (obtained at the corresponding energies in parantheses) and the full figure of merit $ZT$ (including phonon contributions and calculated at the same energy) are listed in the bottom rows. All energies and the coupling strength $|t_{sg}|$ are given in eV, while S is defined in $\mu V/K$, and $ZT_{el}$ and $ZT$ are dimensionless and calculated at the temperature $T=300\,$K. Reprinted with permission from Ref.~\cite{StadlerFano}, Copyright (2011), American Institute of Physics.}
\label{tab.dftresults}
\begin{tabular}{p{2cm}p{2.4cm}p{2cm}p{2.4cm}p{2.4cm}}
\hline\noalign{\smallskip}
  molecule       & C9-CH$_2$-5 & C9-O-4 & C8-NO$_2$-4 & C9-NO$_2$-5 \\
\noalign{\smallskip}\hline\noalign{\smallskip}
$|\Delta\varepsilon|$ & 0.40 &  0.91 & 0.61 & 1.60 \\
$|t_{sg}|$  & 1.03 & 1.12 & 0.32 & 0.40 \\
\noalign{\smallskip}\hline\noalign{\smallskip}
$E_0$  & 0.26 &  -1.44 & 1.61 & 1.72 \\
$|\Delta E|$ & 0.89 & 0.65 & 0.40 & 0.20 \\
\noalign{\smallskip}\hline\noalign{\smallskip}
$S$ ($E$) & -170 (0.21) & 184 (-1.39) & 199 (1.65) & 349 (1.74) \\
$ZT_{el}$ ($E$) & 0.82 (0.19) & 0.93 (-1.37) & 1.00 (1.67) & 3.00 (1.77) \\
$ZT$ ($E$) & 0.003 (0.19) & 0.01 (-1.37) & 0.03 (1.67) & 0.48 (1.77)\\
\noalign{\smallskip}\hline\noalign{\smallskip}
\end{tabular}
\end{table}

Finally, the suitability of four candidates for t-stub molecules for thermoelectric devices is addressed in Table~\ref{tab.dftresults}, where the thermopower $S$ and the figure of merit $ZT$ are explicitly calculated from DFT. The table shows both the pure electronic $ZT_{el}$, which is completely determined by the electronic transmission function assuming a vanishing phonon contribution to the thermal conductance, $\kappa_{ph}$, and the full $ZT$ including a finite $\kappa_{ph}$, for which a constant value of 50 pW/K was used. Notably, the qualitative ranking in $S$ and $ZT$ reversely corresponds to that of $|\Delta E|$, reflecting that the more asymmetric $\mathcal{T}(E)$ curves result in higher values for thermopower and figure of merit. This remains true and is at the highest end even emphasized when the contributions of phonons are considered. The two drawbacks for C9-NO$_2$-5 in terms of applications are the rather high value of $E_0$ and its instability as a radical. These results demonstrate that for an efficient and robust way of implementing QI induced transmission minima for thermoelectric applications an asymmetric transmission function with a transmission peak located close to the minimum and both near E$_F$ in a chemically stable structure is needed. Since the analysis in Table~\ref{tab.dftresults} covers a large part of the chemical possibilities for t-stub type molecules, i.e. simple wires with one side group, one has to conclude that more complex (and probably aromatic) molecules need to be investigated systematically with regard to their suitability for thermoelectric devices.

\section{Summary}
\label{sec:5}
The purpose of this article was to review my involvement in the attempt to use quantum interference effects for the design of memory elements, logic gates and thermoelectric devices in single molecule electronics. In the first half of this review, I recaptured the derivation of a graphical scheme which is based on a simple tight-binding model of the $\pi$ electron system but can be applied to a broad range of realistic molecular junctions. By comparison to first-principles electron transport calculations, it was demonstrated that the scheme correctly predicts the presence/absence of transmission antiresonances at the Fermi energy for ten different configurations of anthraquinone as well as for cross-conjugated molecules. The graphical scheme provides a direct link between molecular structure and QI and should be a useful guideline in the design of molecules with specific transport properties.

The second part of this article focused on thermoelectric applications, where an engineering of the peak shape in the transmission function was suggested for an enhancement of the electronic contribution to the thermopower and figure of merit. Asymmetric peak shapes are highly desirable in terms of device applications, because they promise high thermoelectric efficiency, and the work presented here therefore provides a route towards the chemical engineering of single molecule junctions which can be used as Peltier elements. For this purpose another TB scheme was introduced for the systematic investigation of the dependence of the symmetry of the transmission function around the energy of QI induced minima on the molecular structure. Within this scheme the molecular orbitals of molecular wires with side groups were mapped onto simple two-site TB models where it was found that the shape of the transmission function just depends on two parameters, namely the coupling between the side group and the molecular chain and the energy difference of the side group atomic orbital and the molecular orbital closest to it. The role of the topology of this molecular orbital in producing quite different peak shapes for very similar molecules could be explained from our model and was further verified by DFT based electron transport calculations.

\subsection*{Acknowledgements}
R.S. is currently supported by the Austrian Science Fund FWF, project Nr. P22548, and is deeply indebted to his collaborators in the work reviewed in this article, namely Troels Markussen, Kristian S. Thygesen, Mike Forshaw, Christian Joachim and Stephane Ami.

\end{document}